\documentclass[aps,prb,reprint]{revtex4-1}
\usepackage{graphicx}
\usepackage{array}
\usepackage{float}
\usepackage{textcomp}
\usepackage{graphicx}% Include figure files
\usepackage{dcolumn}% Align table columns on decimal point
\usepackage{bm}% bold math
\usepackage{amsmath}
\usepackage{amsthm}
\usepackage{amsfonts}
\usepackage{color}
\usepackage{float}
\usepackage{hyperref}% add hypertext capabilities
\begin{document}

%You can use \jpcs to insert 'Journal of Physics: Conference Series' in italics
\title{Noise and loss of superconducting aluminium resonators at single photon energies}

%\author{Jonathan~Burnett, Andreas~Bengtsson, David~Niepce \& Jonas~Bylander}
\author{Jonathan~Burnett}
\author{Andreas~Bengtsson}
\author{David~Niepce}
\author{Jonas~Bylander}
%\address{Chalmers University of Technology, Department of Microtechnology and Nanoscience, G\"oteborg, Sweden}
\affiliation{Chalmers University of Technology, Department of Microtechnology and Nanoscience, G\"oteborg, Sweden}

%\ead{burnett@chalmers.se}
\email{burnett@chalmers.se}

\begin{abstract}%you must include an abstract
The loss and noise mechanisms of superconducting resonators are useful tools for understanding decoherence in superconducting circuits. While the loss mechanisms have been heavily studied, noise in superconducting resonators has only recently been investigated. In particular, there is an absence of literature on noise in the single photon limit. Here, we measure the loss and noise of an aluminium on silicon quarter-wavelength ($\lambda/4$) resonator in the single photon regime. 
\end{abstract}

\maketitle

\begin{figure*}
\centering
\includegraphics[width=1\linewidth]{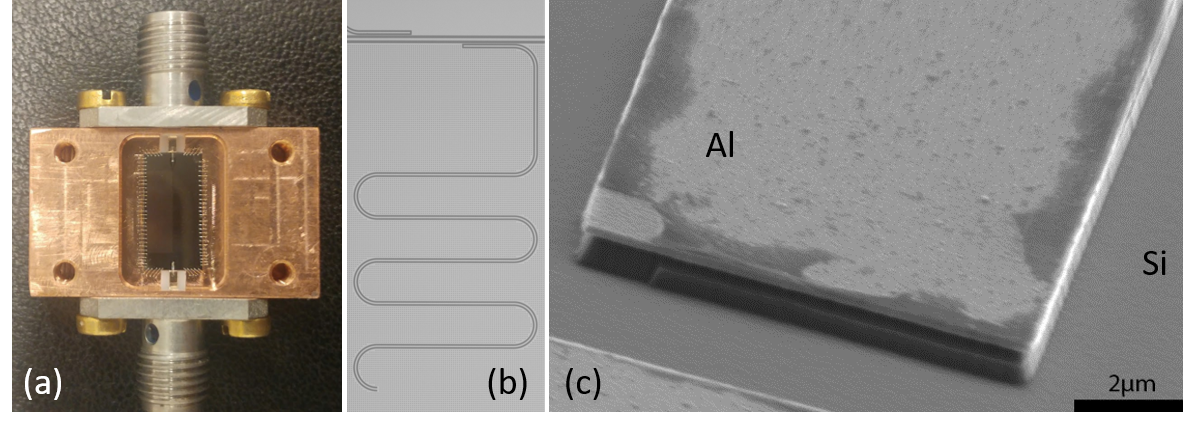}%Resonator_tidy1}
\caption{{\bf (a)} An optical photograph of a copper sample enclosure, with sample in the centre.{\bf (b)} An optical photograph of a $\lambda/4$ resonator, capacitively coupled to a microwave transmission line. The Si substrate is shown by the darker gray tone, while the Al superconductor is shown by the lighter gray tone. The ground plane contains holes for flux trapping. {\bf (c)} A scanning electron micrograph of the open end of the central conductor of the resonator. The image is taken at an angle to demonstrate the trenching of the Si substrate.}
\label{respic1}
\end{figure*}

\begin{figure}
\centering
\includegraphics[width=1\linewidth]{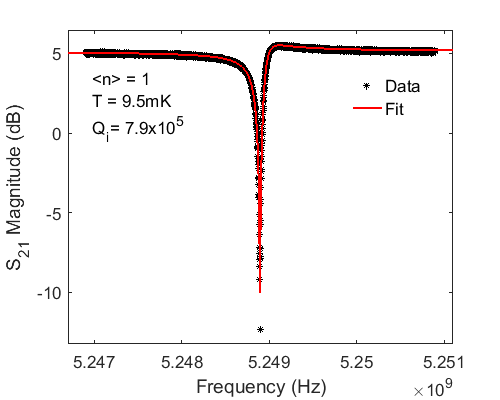}
\caption{Plot of the $S_{21}$ magnitude response of the superconducting resonator. In red is a fit to the data which extracts the resonator parameters. The measurement was performed at 9.5~mK, with $P_{\rm app}$~=~-146~dBm, which corresponds to $\left<n\right>$~=~1.}
\label{resfit1}
\end{figure}

\section{Introduction}

Over the past decade it has become clear that the main causes of decoherence in quantum circuits are bad microwave engineering, excess quasiparticles and parasitic two level systems (TLS). As the demands on quantum circuits increase, the tolerance to decoherence mechanisms decreases. Therefore, despite progress in tackling these issues, they remain the primary problem to overcome when attempting to manufacture large scale quantum systems.

Superconducting resonators have become a common tool in understanding the decoherence mechanisms of quantum circuits. When fabricated from the same materials, using the same processes, the resonator quality factor can provide a good estimate of the qubit relaxation rate \cite{dunsworth2017characterization}. This has motivated studies of the quality factor as a function of material\cite{sage2011study}, deposition\cite{megrant2012planar}, etching\cite{sandbergetch,bruno2015reducing}, magnetic fields\cite{chairofields}, microwave engineering\cite{wennercrosstalk} and quasiparticle trapping\cite{Nsanzinezaquasiparticle}.

A prominent feature of these studies is related to the origin, nature and removal of TLS\cite{muller2017towards,Romanenko2017}. On the nature of TLS, recent results unambiguously demonstrated the the existence of a TLS-TLS interaction\cite{lisenfeld2015observation}. TLS-TLS interactions have been suggested within literature focusing on the low temperature properties of glasses. Specifically, such an interaction was proposed as the source of broadening in resonant absorption measurements in burosilicate glass\cite{ARNOLD1975883} and as a potential explanation in dipole-gap measurements in thin film SiO$_x$ capacitors\cite{Salvinodipolegap,carruzzo,Burin1995}. Within the context of superconducting devices, the interacting nature of TLS is revealed in the temperature dependence of $1/f$ noise\cite{burnett2014evidence,burnett2016analysis,ramanayaka2015evidence,faoro2012internal,faoro2015interacting}. These interactions result in TLS moving in and out of resonance with a superconducting device. This effect has been directly observed in superconducting qubits\cite{mullertlsnoise}, where it leads to a time-varying qubit relaxation rate. Analogously, in superconducting resonators this would produce a time-varying single-photon $Q_{\rm i}$. While previous measurements have measured the TLS-TLS interaction,  there is still no precise measurement of either the TLS-TLS interaction strength or the timescale for the interaction. In principle, these would be most directly found by performing noise measurements at single photon energies. However, to date noise studies have proved non trivial to perform for fewer than 10 photons in the resonator\cite{burnett2014evidence,burnett2016analysis,ramanayaka2015evidence,neill2013fluctuations}.

Here, we study the loss and noise of a superconducting aluminium $\lambda/4$ resonator. We demonstrate a level of loss which is comparable to the literature. We then study the noise of this resonator at single photon energies. This opens up the possibility of directly measuring noise in superconducting qubits as well as further examining the nature of interacting TLS which are the limiting factor for many quantum circuits.

\section{Sample details}

Fabrication of the device begins with a solvent clean of a high resistivity silicon wafer. Following this, the wafer is submerged in a 2\% hydrofluoric acid bath to remove the native surface oxide and passivate the surface with hydrogen. Within 3 minutes, the wafer is placed under vacuum inside the load lock of a Plassys MEB deposition system. The wafer is then heated to 300$^o$C while the vacuum chamber pumps. Once the wafer has cooled to room temperature and a base pressure of 1.1x10$^{-7}$~mbar is reached, 150~nm of Al is deposited at a rate of 0.5~nm/s. Next, the vacuum chamber is filled to 10~mbar of 99.99\% pure molecular oxygen for 10 minutes, after this the chamber vented to atmosphere. A 1.2~$\mu$m thick layer of AZ1512HS photoresist is then patterned by direct-write laser lithography to realise the microwave circuitry. The photoresist is developed in AZ developer diluted with H$_2$O 1:1, which minimises the parasitic etching of aluminium. This pattern is transferred into the Al film by a wet etch in a mixture of phosphoric, nitric, and acetic acids. Then, a reactive ion etch using an inductively coupled NF$_3$ plasma was used to isotropically etch the Si substrate, forming a 1~$\mu$m deep trench with a 400~nm undercut below the Al features. After dicing, the resulting chip is cleaned using hot solvents, then wirebonded within a light-tight connectorised copper sample enclosure (shown in Fig.~\ref{respic1}a). This sample enclosure is then placed on a gold-plated copper cold finger at the 9.5~mK stage of a dilution refrigerator. A photograph of a typical microwave resonator, and a SEM image of an etched area, are shown in Fig.~\ref{respic1} b and c, respectively. The black residues close to the aluminium edge are indicative of burnt resist from the RIE process. This is supported by samples that only had a wet etch not showing these residues.

\section{Dielectric loss measurements}

\begin{figure}
\centering
\includegraphics[width=1\linewidth]{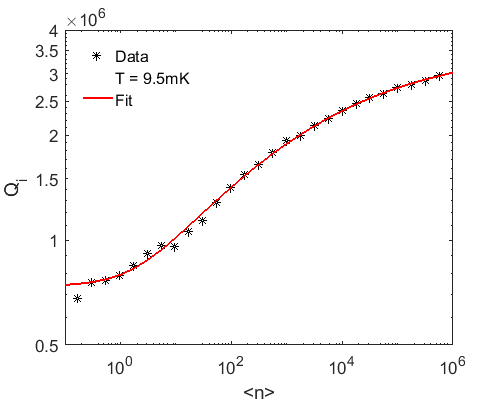}
\caption{Plot of $Q_{\rm i}$ as a function of $\left<n\right>$ for the resonator at 9.5~mK. In red is a fit to TLS losses described by Eq.~\eqref{losseq1}.}
\label{lossfit1}
\end{figure}

The $S_{21}$ transmission response of the superconducting resonator is measured at 9.5~mK while the microwave power is varied. A traceable fit routine\cite{probst2015efficient} is used to extract the resonant frequency ($\omega_0$), internal quality factor ($Q_{\rm i}$) and coupling quality factor ($Q_{\rm c}$). Figure~\ref{resfit1} shows the fitted $S_{21}$ magnitude response of the resonator at 9.5~mK. For an applied microwave power ($P_{\rm app}$) of -146~dBm, this reveals $\omega_0/2\pi$~=~5.24889~GHz, $Q_{\rm c}$~=~33$\times$10$^3$ and $Q_{\rm i}$~=~7.9$\times$10$^5$. As $P_{\rm app}$ is changed, $Q_{\rm i}$ is found to vary due to depolarisation of TLS\cite{sage2011study,megrant2012planar,sandbergetch,bruno2015reducing,burnett2014evidence,burnett2016analysis,ramanayaka2015evidence}. The average number of photons within the resonator ($\left<n\right>$) varies with $P_{\rm app}$ as\cite{bruno2015reducing} 
\begin{equation}
   \left<n\right> = \frac{\left<E_{\rm int}\right>}{\hbar\omega_0} = \frac{2}{\hbar\omega_0^2}\frac{Z_0}{Z_{\rm r}}\frac{Q_{\rm l}^2}{Q_{\rm c}}P_{\rm app}
\end{equation}
where $Z_0$ is the characteristic impedance ($Z_0$~=~50$\Omega$), $Z_r$ is the resonator impedance ($Z_r$ is chosen to be close to 50$\Omega$), $\left<E_{\rm int}\right>$ is the average energy stored within the resonator and $Q_{\rm l}$ is the loaded quality factor ($1/Q_{\rm l} = 1/Q_{\rm c} + 1/Q_{\rm i}$). Figure~\ref{lossfit1} shows the effect of TLS depolarisation in a measurement of $Q_{\rm i}$ as a function of $\left<n\right>$. This TLS depolarisation can be fit to a TLS-based loss model\cite{burnett2016analysis}

\begin{equation}
    \frac{1}{Q_{\rm i}} = F\delta_{\rm TLS}^{0}\frac{{\rm tanh}\left(\frac{\hbar\omega_0}{2k_{B}T}\right)}{\left(1+\left(\frac{\left<n\right>}{n_c}\right)\right)^\beta} + \delta_{\rm other}
    \label{losseq1}
\end{equation}
where $F$ is the filling factor describing the ratio of $E$-field in the TLS host volume to the total volume. $\delta_{\rm TLS}^{0} = 1/Q_{\rm TLS}$ is the TLS loss tangent, $n_c$ is the critical number of photons within the resonator to generate the $E$-field required to saturate one TLS. $\delta_{\rm other}$ is the contribution from non-TLS loss mechanisms, which are generally associated with loss at high microwave drives. A fit to Eq.~\eqref{losseq1} is shown in Fig.~\ref{lossfit1}. This shows $\delta_{\rm other}~\approx$~1/(3.5$\times$10$^6$), while indicating that $F\delta_{\rm TLS}^{0}\approx~$1/(8.7$\times$10$^5$). Both of these loss rates are comparable with those found for the best Al resonators\cite{megrant2012planar}. Further improvements to $\delta_{\rm other}$ are possible by improving the infrared filtering\cite{barends2011minimizing} or the magnetic screening\cite{chairofields}. While further improvements to $\delta_{\rm TLS}^0$ are also possible with either increased trenching of the substrate\cite{sandbergetch,bruno2015reducing} or by cleaning of resist residues\cite{quintana2014characterization} which are found in Fig.~\ref{respic1}c.

\section{Noise measurements}

The loss induced by the TLS corresponds to the resonant absorption of microwave photons. Recent experiments have examined the role of these resonant TLS in contributing noise to the resonator\cite{burnett2014evidence,neill2013fluctuations}. The dependence of TLS-induced noise has been measured as a function of $P_{\rm app}$ and temperature. Within these studies, the temperature dependence has been thoroughly examined in the range of 50--700~mK. However, the span of $P_{\rm app}$ that was examined corresponds to $\left<n\right>$~=~7--10$^4$. Consequently, the TLS noise in the limit of single-photon excitation has not been examined. Not only is this limit most relevant to dephasing in superconducting qubits, but it is also relevant to revealing properties of TLS in general. 

\begin{figure}
\centering
\includegraphics[width=1\linewidth]{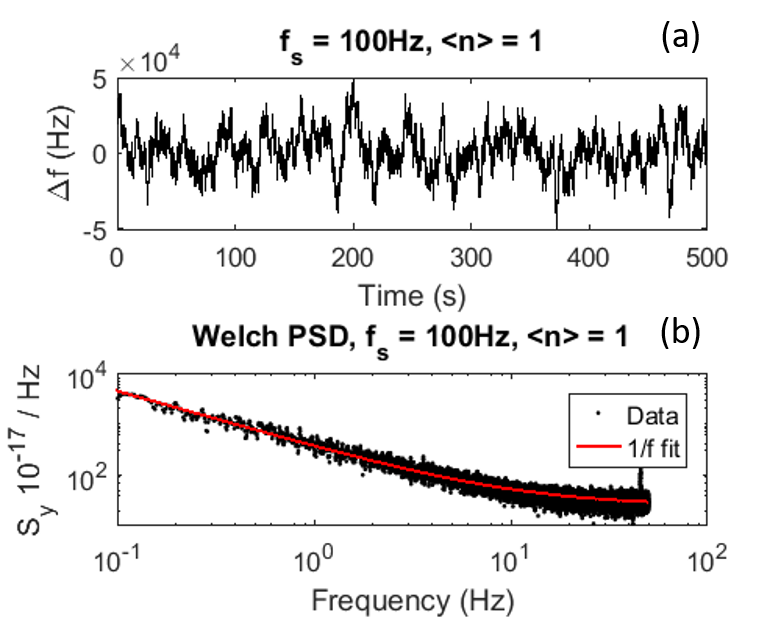}
\caption{{\bf (a)} A plot of the raw frequency jitter of the resonator measured at $\left<n\right>$~=~1 and at a temperature of 9.5~mK. {\bf (b)} A plot of the spectrum of fractional frequency fluctuations ($S_Y$). In red is a fit to a $1/f$ noise model shown in Eq.~\eqref{1fnoisemodel}.}
\label{noiseplot}
\end{figure}

A Pound setup is used to form a frequency locked loop which can continuously monitor $\omega_0(t)$. The Pound setup we use is identical to one previously used in the study of low frequency noise in superconducting resonators\cite{lindstrom2011pound,burnett2013slow,burnett2014evidence}.  Here, the Pound setup is operated with a low bandwidth of 300~Hz. This improves the signal-to-noise, enabling the Pound setup to monitor a resonator at single photon energies. However, the low bandwidth means that this setup is only suitable for studying `slow' fluctuations.  Figure~\ref{noiseplot}a shows a 500~s window of a measurement of the frequency jitter of the resonator, measured at $\left<n\right>$~=~1 and 9.5~mK. This frequency jitter can be better understood by examining the spectrum of frequency fluctuations ($S_{y}$). This is obtained using the Welch spectral density estimate with a 50\% overlap and a Hanning window. The resulting plot of $S_{y}$ is shown in Fig.~\ref{noiseplot}b; this data is fit to a general noise model
\begin{equation}
  S_{y}(f) = \frac{h_{-1}}{f^\alpha} + h_0
  \label{1fnoisemodel}
\end{equation}
where $h_0$ is a white frequency noise level, $h_{-1}$ is a flicker frequency noise level and $\alpha$ is an exponent describing the strength of low frequency noise components. When $\alpha$~=~1, the first term represents a true flicker noise process. From this fit we find that the white noise level is described by $h_0$~=~2.5$\times$10$^{-16}$, while the flicker noise level is described by $\alpha$~=~1.05 and $h_{-1}$~=~3.5$\times$10$^{-15}$. This level of noise is larger than that previously observed\cite{burnett2014evidence,burnett2016analysis,ramanayaka2015evidence}. However, since we measure at both lower microwave powers and at lower temperatures, this is expected from the strong power and temperature dependence of dielectric noise\cite{faoro2015interacting}.

\section{Summary}

In summary, we measured frequency noise of a high-Q superconducting resonator at single photon energies. This is an important step towards studies of the interacting nature of TLS, which are currently limiting the performance of many superconducting circuits. In showing that noise can be measured at single photon energies, the technique could be extended to the circuit-QED architecture. Here, a superconducting qubit in the dispersive regime shifts the resonator frequency with $\chi=-g^2/\Delta$, where $g$ is the qubit--resonator coupling and $\Delta$ is the frequency detuning between the qubit and the resonator. The frequency shift implies that any noise of the qubit frequency will get mapped to a frequency noise of the resonator. For normal values of $g$ and $\Delta$ the effective noise of the resonator will be between $0.01$ and $0.001$ times than that of the qubit. This gives a straightforward method to measure flux noise of qubits by measuring the frequency noise of the resonator, without needing to use advanced pulse sequences\cite{bylander2011noise,yanfluxnoise}. It is important to point out that such measurement should be performed at sub single photon levels in the resonator, to avoid additional noise due to the AC-stark effect \cite{schuster2007resolving}. To have a stable locking of the Pound loop at such low energies, a parametric amplifier between the sample and the semiconductor amplifier, would be needed. We believe that this technique could lead to more efficient investigations of the origins of flux noise in superconducting circuits \cite{kumar2016origin,de2017direct,quintana2017observation}.

%\section{Example of a Table}

%\begin{center}
%\begin{table}[h]
%\caption{caption goes here}
%\centering
%\begin{tabular}{@{}l*{7}{l}}
%\br
%left title&centre title&right title\\
%%\mr
%entry 1&entry 2&entry 3\\
%entry 4&entry 5&entry 6\\
%\br
%\end{tabular}
%\end{table}
%\end{center}

%\subsection{Numbered list example}
%\begin{enumerate}
%\item First item
%\item Second item
%\item Third item
%\item Fourth item
%\item Fifth item
%\end{enumerate}

%\ack%this is an unnumbered acknowledgement section
\begin{acknowledgements}
We acknowledge fruitful discussions with Andrey Danilov and the mechanical work from Lars J\"onsson. Financial support came from the Knut and Alice Wallenberg foundation, and the Swedish research council.
\end{acknowledgements}

\bibliographystyle{h-physrev5}
\bibliography{sust}

\end{document}